\documentstyle[12pt]{article}
\begin{document}
\begin{flushright}
IHEP 95-82\\
hep-ph/9507243\\
\end{flushright}
\vspace*{4mm}
\begin{center}
{\large \bf Hidden scale dependence in renormalon}\\
\vspace*{4mm}
V.V.Kiselev\\
Institute for High Energy Physics,\\
Protvino, Moscow Region, 142284, Russia.\\
E-mail: kiselev@mx.ihep.su\\
Fax: +7-(095)-230 23 37.
\end{center}
\begin{abstract}
In the framework of renormalon consideration, a role of the anomalous
dependence of gluon propagator on the scale $\mu$ is shown for a large number
of quark flavours as well as in the nonabelian theory, where the parameter of
the "running" coupling constant evolution has a hidden dependence on the scale.
A phenomenological matching between the modified renormalon and the static
quark potential is described to fix the renormalon scales.
\end{abstract}

\section*{Introduction}

In the QCD perturbation theory, an account for the next-to-leading order
term over $\alpha_s$ is connected with the uncertainty, caused by a choice
of the energy scale, determining the $\alpha_s(\mu^2)$ value, since the
transition to the $\bar \mu$ scale leads to the substitution
$$
\alpha_s(\mu^2) \approx \alpha_s(\bar \mu^2)\biggl(1 - \alpha_s(\bar \mu^2)
\frac{\beta_0}{4\pi} \ln\frac{\mu^2}{\bar \mu^2}\biggr)\; ,
$$
where $\beta_0 =(11 N - 2n_f)/3$, $N$ is the number of colours, $n_f$ is
the number of quark flavours. So, the physical quantity, represented in the
form
$$
r = r_0 \alpha_s(\mu^2)\{1+ r_1 \alpha_s + O(\alpha_s^2)\}
$$
with the given order of accuracy, will get the form
\begin{eqnarray}
r & = & r_0 \alpha_s(\bar \mu^2)\bigg\{1+ \biggl(r_1 - \frac{\beta_0}{4\pi}
\ln\frac{\mu^2}{\bar \mu^2}\biggr) \alpha_s + O(\alpha_s^2)\bigg\} \nonumber\\
&=& r_0 \alpha_s(\bar \mu^2)\{1+\bar r_1 \alpha_s + O(\alpha_s^2)\}\;.\nonumber
\end{eqnarray}
Thus, the value of $r_1$ coefficient for the $\alpha_s$ correction depends on
the scale of $\alpha_s(\mu^2)$ determination. In ref.\cite{blm}, one
offered to fix the $\mu$ scale so that the $r_1$ coefficient does not
contain terms, proportional to $\beta_0$. The latter procedure can be
understood as one, leading to $r_1$ to be independent of the number of
quark flavours $n_f \sim \beta_0$. As was shown in refs.\cite{2,3}, such
choice of scale has the strict sense in the framework of the $1/n_f$ expansion,
where the $\alpha_s$ correction to the gluon propagator is determined by
the fermion loop contribution into the vacuum polarization. So, the expression
for such contribution depends on the regularization scheme, and in the
next-to-leading order one has
\begin{equation}
\alpha_s(\mu^2) D(k^2, \mu) = \frac{\alpha_s(\mu^2)}{k^2}\;
\bigg\{1+ \frac{\alpha_s n_f}{6\pi}
\biggl(\ln\frac{k^2}{\mu^2} + C\biggr)\bigg\}\;,
\label{1}
\end{equation}
where $D(k^2,\mu)$ determines the transversal part of the gluon propagator
$D^{ab}_{\nu\lambda}(k^2,\mu) = \delta^{ab} D(k^2,\mu) (-g_{\nu\lambda}+
k_\nu k_\lambda/k^2)$. The constant value $C$ is defined by the renormalization
scheme, so $C_{\overline{\rm MS}} = -5/3$ and $C_V =0$ in the
so-called $V$-scheme \cite{blm}.

The summing of $(n_f\alpha_s)^n$ contributions into the gluon propagator
leads to the expression
\begin{equation}
\alpha_s(\mu^2) D(k^2, \mu) = \frac{\alpha_s(e^C k^2)}{k^2}\;,
\label{g1}
\end{equation}
i.e., in fact, it results in the account for the "running" $\alpha_s$ value in
respect to the gluon virtuality. Note, the $1/n_f$ consideration is exact
for the abelian theory, where $\alpha_s D$ is the renormalization group
invariant. Moreover, the $e^{-C}\Lambda^2_{QCD}$ value does not depend on the
renormalization scheme, and therefore it results in the scheme-independence
for the expression
$$
\alpha_s(e^C k^2) = \frac{4\pi}{\beta_0\ln(e^C k^2/\Lambda^2_{QCD})}\;.
$$

As one can see in the performed consideration, the transition to the
nonabelian theory is done by the $2n_f/3 \to -\beta_0$ substitution, that is
called as the procedure of "naive nonabelization", giving the correct
"running" of the coupling constant.

Further, consider the physical quantity $r$, for which the first order
$\alpha_s$-contribution is calculated as an integral over the gluon
virtuality with the weight $F(k^2)$
$$
r = \int \frac{dk^2}{k^2}\; \alpha_s(\mu^2) F(k^2)\;.
$$
Then the account for the $(n_f\alpha_s)^n$ corrections to the gluon propagator
leads to the substitution of "running" value $\alpha_s(e^C k^2)$ for
$\alpha_s(\mu^2)$
\begin{equation}
\alpha_s(\mu^2) \to \alpha_s(\mu^2)\sum_{n=0}^{\infty}
\bigg\{-\frac{\alpha_s(\mu^2)\beta_0}{4\pi}\biggl(\ln\frac{k^2}{\mu^2}+C\biggr)
\bigg\}^n\;,
\label{2}
\end{equation}
so that in the offered procedure of the scale fixing \cite{blm}, the
introduction of the first order $n_f\alpha_s$-correction results in the
substitution
$\mu^2\to \bar \mu^2 = \mu^2 \exp\{C+\langle\ln k^2/\mu^2\rangle\}$, where
the average value is determined by the integral with the $F$ weight.
However, in some cases \cite{2,3} the integration of the $n$-th order term
in expansion (\ref{2}) leads to the $n!$ factorial growth of the coefficients
for the expansion over $\alpha_s^n(\mu^2)$, so that this rising can be
determined by the region of low virtualities as well as large ones of gluon.
Then one says about the infrared and ultraviolet renormalons, respectively
\cite{2,3,4}. As was shown \cite{2,3}, the renormalon leads to power-like
uncertainties of the $r$ evaluation, so
\begin{equation}
\Delta r = \biggl(\frac{\Lambda_{QCD}}{\mu}\biggr)^k a_k\;,
\label{3}
\end{equation}
where $k$ is positive for the infrared renormalon and it is negative for the
ultraviolet one. Indeed, consider a quantity ${\cal F}$, presented in the form
\begin{equation}
{\cal F}(\alpha_s) = \sum_{n=0}^{\infty} f_n \biggl(\frac{\beta_0}{4\pi}
\alpha_s(\mu^2)\biggr)^n\;, \label{f}
\end{equation}
and its Borel transformation, defined as
$$
B[{\cal F}](u) = f_0\delta(u)+
\sum_{n=0}^{\infty} \frac{f_{n+1}}{n!} u^n\;,
$$
so that ${\cal F}$ can be restored due to the integral operation
\begin{equation}
{\cal F}(\alpha_s) = \int_0^\infty du B[{\cal F}](u) \exp
\biggl(- \frac{4\pi u}{\beta_0\alpha_s(\mu^2)}\biggr)\;.\label{if}
\end{equation}
In the case of the factorial divergence in the coefficients of expansion
(\ref{f}), the Borel transformation has singular points over $u$ \cite{2,3},
so that after transform (\ref{if}), a pole contribution at a point
$u_k=k/2$
$$
\Delta B[{\cal F}](u) = \frac{a_k}{u-u_k}
$$
depends on the rule of a way out the singularity, and
$$
\Delta {\cal F}\sim a_k \exp
\biggl(- \frac{4\pi u_k}{\beta_0\alpha_s(\mu^2)}\biggr)=
a_k \biggl(\frac{\mu}{\Lambda_{QCD}}\biggr)^{-2u_k} =
a_k\biggl(\frac{\Lambda_{QCD}}{\mu}\biggr)^k\;.
$$

For the two-point correlator of heavy quark vector currents,
for instance, one has $k=4$ and the corresponding uncertainty can be, in fact,
eliminated in the procedure of definition for the nonperturbative gluon
condensate, having the same power over the infrared parameter \cite{3}.
However,  for the renormalized mass of heavy quark the infrared renormalon
with $k=1$ does not correspond to some definite quark-gluon condensate, and
the value $\Delta m(\mu) \sim \Lambda_{QCD}$ can not be adopted into a
definition of a physical condensate.

In the present paper we modify the renormalon in the framework of
QCD one-loop renormalization group and in the $1/n_f$ consideration
by the account for the anomalous dependence of the gluon propagator on the
$\mu$ scale, so that under the "naive nonabelization", the modification appears
in the determination of the evolution parameter for the "running" coupling
constant in different schemes of the renormalization. The standard renormalon
corresponds to the specific choice of the fixed point of the gluon
propagator dependence on the scale. The nonabelian evolution of the gluon
propagator leads to a hidden scale dependence in the renormalon, because of
the anomalous dependence on the $\mu$ scale. A phenomenological matching
of the renormalization group invariant parameters of the modified renormalon
$\Lambda_{QCD}$ and $\mu_g$, corresponding to the singularity points in the QCD
coupling constant and in the gluon propagator, respectively,
with the tension of string, giving the linear rise of the potential in the
heavy quarkonium, and with the constant of the subleading coulomb interaction
allows one to fix the physical parameters, which, in such way, point to a
deviation from the standard renormalon on the 1$\sigma$ confidence level.
At asymptotically high virtualities of gluon, the presence of the second scale
in addition to the evolution parameter of the QCD "running" coupling constant
is not essential. However, near the infrared region, the presence of the
additional scale has to be included into the consideration, as it has for
the potential in the heavy quarkonium or, for instance, for a construction of
models for such nonperturbative quantities as the gluon condensate.

In Section 1 we use the one-loop renormalization group of QCD to find the
anomalous scale dependence of the gluon propagator. In Section 2 we modify
the renormalon in the framework of $1/n_f$ consideration and under the
procedure of "naive nonabelization" for the "running" constant. In
Section 3 we find the hidden scale dependence in the renormalon, because of the
nonabelian evolution of the gluon propagator. In Section 4 we make the
phenomenological matching of the modified renormalon parameters with the
potential in the heavy quarkonium. In Section 5 we use a toy ansatz, motivated
by the modified renormalon, to make a model estimate of the gluon
condensate value. In Conclusion the obtained results are summarized.

\section{Anomalous dependence of gluon propagator on scale}

In the covariant gauge, the gluon propagator has the form
$$
D^{ab}_{\nu\lambda}(k^2,\mu) = \frac{\delta^{ab}}{k^2\omega(\mu^2,k^2)}
\biggl(-g_{\nu\lambda}+(1-a_l(\mu^2)\omega(\mu^2,k^2))
\frac{k_\nu k_\lambda}{k^2}\biggr)
\;,
$$
where $a_l\omega$ is the renormalization group invariant, since the
renormalization group transformations make "bare" quantities, independent of
the scale, in the form
\begin{eqnarray}
a_l^B & = & Z_3 a_l\;,\nonumber\\
\omega^B & = & Z_3^{-1} \omega\;,\nonumber
\end{eqnarray}
where $Z_3$ is the renormalization constant of the gluon field $A_\mu^B=
Z_3^{1/2} A_\mu$. In the MS-scheme, one-loop calculations give
$$
Z_3^{\rm MS} = 1+\frac{1}{\epsilon}\;
\frac{\alpha_s}{24\pi}(N (13-3a_l)-4n_f)\;.
$$
$Z_3^{\overline{\rm MS}}$ can be obtained from $Z_3^{\rm MS}$ by the
substitution $1/\epsilon\to 1/\epsilon +\ln 4\pi-\gamma_E$, where
$\gamma_E=0.5772\ldots$ The differential equation for $a_l$ has the
following $\overline{\rm MS}$-scheme form
\begin{equation}
\frac{da_l(\tau)}{d\tau} = \frac{\alpha_s}{12\pi}\; a_l(\tau)
[(13N-4n_f)-3Na_l(\tau)]\;, \label{eal}
\end{equation}
where $\tau=\ln(\mu/\Lambda_{QCD})$, and $\alpha_s(\tau) = 2\pi/(\beta_0\tau)$.
Eq.(\ref{eal}) has the solution
\begin{equation}
a_l(\tau) = \frac{(13N-4n_f)C_a\tau^{n_a}}{1+3NC_a\tau^{n_a}}\;,\label{sol}
\end{equation}
where
$$n_a =\frac{13N-4n_f}{6\beta_0}\;,$$
and $C_a$ is the renormalization group invariant.

Then eq.(\ref{sol}) allows one to represent the gluon propagator in the form
\cite{kis}
\begin{equation}
D^{ab}_{\nu\lambda}(k^2,\mu) = \frac{\delta^{ab}}{k^2}
\frac{1}{1-(1-\omega(\mu_0^2))\biggl(\frac{\alpha_s(\mu^2)}
{\alpha_s(\mu_0^2)}\biggr)^{n_a}}
\biggl(-g_{\nu\lambda}+(1-b)
\frac{k_\nu k_\lambda}{k^2}\biggr)\;,
\label{4}
\end{equation}
where $b$ is the arbitrary gauge
parameter\footnote{In ref.\cite{kis} the solution with the fixed $b=3$ value is
considered and $\omega$ is expressed through $b/a_l$.}, being the
renormalization group invariant.
In eq.(\ref{4}) we take into account the anomalous dependence for the gluon
propagator on the scale, $\omega=\omega(\mu^2)$, only.

The choice of fixed point $\omega \equiv 1$ corresponds to the standard
renormalon. The consideration of the gluon propagator at $\omega\neq 1$
leads to the modified renormalon.

\section{Modified renormalon in $1/n_f$ consideration}

In the leading order over $1/n_f$ one has $n_a=1$, and the cases with
1) $\omega(\mu_0^2)>1$, 2) $\omega(\mu_0^2)=1$ and 3) $\omega(\mu_0^2)<1$
at some large $\mu_0$ values can be formally come to the different choices
of the renormalization group invariant $\mu_g$, such that $\omega(\mu_g^2)=0$
and 1) $\mu_g<\Lambda_{QCD}$, 2)  $\mu_g=\Lambda_{QCD}$ and 3)
$\mu_g>\Lambda_{QCD}$, respectively. The $\alpha_s(\mu_g^2)$ value
is considered as the formal one-loop expression
$$
\alpha_s(\mu^2) = \frac{4\pi}{\beta_0\ln\frac{\mu^2}{\Lambda^2_{QCD}}}\;.
$$
Thus, the $1/n_f$ consideration gives
\begin{equation}
\frac{\alpha_s(\mu^2)}{\omega(\mu^2)} = \frac{\alpha_s(\mu^2)
\alpha_s(\mu_g^2)}{\alpha_s(\mu_g^2) - \alpha_s(\mu^2)} =
\alpha_s(e^d\mu^2)\;,
\label{5}
\end{equation}
where $d$ is the scheme-independent invariant of the one-loop renormalization
group, and it is equal to
$$
d=\ln\frac{\Lambda^2_{QCD}}{\mu_g^2}=
-\frac{4\pi}{\beta_0 \alpha_s(\mu_g^2)}\;.
$$
Further, one can repeat the calculation of $(n_f\alpha_s)^n$ contributions
into the gluon propagator with the substitution of value (\ref{5}) for
$\alpha_s(\mu^2)$, so that this replacement corresponds to the account for
the anomalous dependence of the gluon propagator versus the $\mu$ scale.
Then the scale fixing due to the next-to-leading order $\alpha_s$ correction
results in the expression
$$
\bar \mu^2 = \mu^2 \exp\{C+\langle\ln k^2/\mu^2\rangle+d\}\;,
$$
and the summing of the corresponding contributions modifies eq.(\ref{g1})
\begin{equation}
\alpha_s(\mu^2) D(k^2, \mu) = \frac{\alpha_s(e^{C+d} k^2)}{k^2}=
\frac{\alpha_s(e^C k^2)}{\omega(e^C k^2)k^2}\;.
\label{g2}
\end{equation}

Next, one can define the $\overline{V}$-scheme, introducing
$C_{\overline{V}}=-d$. Then
$$
\Lambda^{\overline{V}}_{QCD} = e^{5/6-d/2}\Lambda^{\overline{\rm MS}}_{QCD}\;.
$$
In the $\overline{V}$-scheme the perturbative potential between the
heavy quark and antiquark in the colour-singlet state will have the form
\begin{equation}
V(q^2) = - \frac{4}{3}\; \frac{4\pi\alpha_s^{\overline{V}}(q^2)}{q^2}
\label{v}
\end{equation}
at $q^2 = {\bf k}^2$. Potential (\ref{v}) comes to the Richardson's potential
\cite{rich}, when one uses $\alpha_s^{\overline{V}}(q^2+\Lambda^2)$
instead of $\alpha_s^{\overline{V}}(q^2)$, and $\Lambda$ fixes the
linearly rising part of potential, confining quarks with the distance increase,
\begin{equation}
\Delta V_{lin}({\bf x}) = \sigma |{\bf x}| =
\frac{8\pi}{27}\; \Lambda^2_{lin} |{\bf x}|\;.\label{8-}
\end{equation}
The fitting of mass spectra for the charmonium and bottomonium in the
Richardson's potential gives
$$
\Lambda_{lin} = 398\; \mbox{MeV}\;.
$$
However, one usually takes $\sigma= 0.18\pm 0.01$ GeV$^2$ as the scale
normalization for the lattice computations, so that this string tension is
equal to that of the Cornell model \cite{cor}. Therefore, we will use
\begin{equation}
\Lambda_{lin} = 425\pm 15\; \mbox{MeV}\;.\label{vbar}
\end{equation}
Value (\ref{vbar}) has to be compared with the evolution scale $\Lambda$ of
the "running" coupling constant of QCD, where
$\alpha_s^{\overline{\rm MS}}(m_Z^2) = 0.117\pm 0.005$ corresponds to
the one-loop value $\Lambda^{\overline{\rm MS}}_{QCD} = 85\pm 25\; \mbox{MeV}$
\cite{pdg}. If one assumes, that
$\Lambda$ does not depend on the number of flavours, then at the
scale of "switching on (off)" the additional flavour of quarks $\mu=m_{n_f+1}$,
the QCD coupling has a discontinuity, related with the step-like change
$\beta_0(n_f)\to \beta_0(n_f+1)=\beta_0(n_f)-2/3$. One can avoid such
discontinuities, if one supposes that $\Lambda$ depends on the
flavour number, so that $\alpha_s(\mu^2=m^2_{n_f+1},\Lambda^{(n_f)},
n_f)=\alpha_s(\mu^2=m^2_{n_f+1},\Lambda^{(n_f+1)},n_f+1)$.
Then in the one-loop approximation one finds
\begin{equation}
\Lambda^{(n_f)} = \Lambda^{(n_f+1)}\biggl(\frac{m_{n_f+1}}
{\Lambda^{(n_f+1)}}\biggr)^\frac{2}{3\beta_0(n_f)}\;.
\label{l}
\end{equation}
Setting $\Lambda^{(5)}=85\pm 25$ MeV, one gets
\begin{equation}
e^{-d/2} = 1.42\pm 0.40\;. \label{e2}
\end{equation}
Hence, the $\alpha_s$ rescaling, accounting for eq.(\ref{l}), results in
the $d$ value, that agrees with the standard renormalon ($d\equiv 0$)
within the current accuracy up to $1\sigma$, corresponding to the
confidence probability, equal to 30\%.

However, the quantity $\alpha_s D$, obtained in the framework of $1/n_f$
consideration under the procedure of substitution $2n_f/3\to -\beta_0(n_f)$,
possesses the renormalization group properties, which disagree with the scale
dependence of $\alpha_s D$ in the nonabelian theory, where $n_a\neq 1$. The
restoring of correct properties in the nonabelian renormalization group
for the gluon propagator results in a hidden scale dependence of the
renormalon.

\section{Nonabelian generalization and hidden scale dependence}

One has $n_a\neq 1$ in the nonabelian theory. Introduce the effective
coupling constant $\bar \alpha_s$
$$
\bar \alpha_s(\mu^2,\Lambda(\mu)) = \frac{\alpha_s(\mu^2)}{\omega(\mu^2)} =
\frac{4\pi}{\beta_0\ln(\mu^2/\Lambda^2(\mu))}\;,
$$
where the evolution parameter $\Lambda(\mu)$ has the following dependence
on the $\mu$ scale
\begin{equation}
\Lambda(\mu) = \Lambda_{QCD}\biggl(\frac{\mu}{\Lambda_{QCD}}\biggr)^{\biggl(
\frac{\ln(\mu_g/\Lambda_{QCD})}{\ln(\mu/\Lambda_{QCD})}\biggr)^{n_a}}\;.
\label{h}
\end{equation}
Thus, expansions in the perturbative theory of QCD built over the powers of
$\bar \alpha_s$, that has the following transformation properties, caused by
the hidden scale dependence,
\begin{eqnarray}
\bar \alpha_s(\mu_2^2,\Lambda(\mu)) = \frac{\bar \alpha_s(\mu_1^2,
\Lambda(\mu))}{1+\frac{\beta_0}{4\pi}\bar \alpha_s(\mu_1^2,\Lambda(\mu))\ln
\frac{\mu_2^2}{\mu_1^2}}\;,\label{t1}\\
\bar \alpha_s(\mu_2^2,\Lambda(\mu_2)) = \frac{\bar \alpha_s(\mu_1^2,
\Lambda(\mu_1))}{1+\frac{\beta_0}{4\pi}\bar \alpha_s(\mu_1^2,\Lambda(\mu_1))\ln
\frac{\mu_2^2}{\mu_1^2}}\;F(\mu_2,\mu_1)\;,\label{t2}
\end{eqnarray}
where
$$
F(\mu_2,\mu_1) = \frac
{1-(1-\omega(\mu_1^2))(\alpha_s(\mu_2^2)/\alpha_s(\mu_1^2))}
{1-(1-\omega(\mu_1^2))(\alpha_s(\mu_2^2)/\alpha_s(\mu_1^2))^{n_a}}\;.
$$
One finds $F\equiv 1$ at $\omega \equiv 1$ or $n_a=1$, so that the choice
of fixed point or the abelian case make us to return to the consideration,
performed in the previous section. Then $\Lambda(\mu) = \mu_g$, i.e. we come to
the $\overline{V}$-scheme, and the standard consideration with
$\mu_g= \Lambda_{QCD}$ becomes valid for the choice of fixed point.

Further, it is evident, that the introduction of the fermion loop contribution
into the gluon propagator results in the expression
\begin{equation}
\alpha_s(\mu^2) D(k^2, \mu) = \frac{\bar \alpha_s(e^C k^2,\Lambda(\mu))}{k^2} =
\frac{\alpha_s(e^{C+d(\mu)} k^2)}{k^2}\;,\label{em}
\end{equation}
where the hidden dependence on the $\mu$ scale in $\Lambda(\mu)$ or in
$d(\mu)$ restores the correct renormalization group properties of the
$\alpha_s D$ quantity in the nonabelian theory, so that
$$
d(\mu)=\ln\frac{\Lambda^2_{QCD}}{\Lambda^2(\mu)}\;.
$$
Accounting for eq.(\ref{em}), one can easily get the generalization of
equations, giving the BLM-procedure of scale fixing in the "running" constant
and the renormalon modification under $d\to d(\mu)$.

However, physically measurable quantities have to be invariant under the
renormalization group transformations, so that corresponding expressions can
contain the $\alpha_s D$ value, given at a physical scale, being invariant
under the renormalization group and defining a normalization condition.

\section{Phenomenology of scale fixing}

Let us consider the momentum-space potential in the heavy quarkonium at
$q^2\to 0$ in the $V$-scheme under the Richardson's prescription
$$
V(q^2) =  - \frac{4}{3}\; \frac{4\pi}{q^2}\; \frac{\alpha_s(\mu^2(q^2))}
{\omega(\mu^2(q^2))}\;,
$$
where $\mu^2(q^2)= q^2+\mu^2_g$, so that the expansion over the $q^2/\mu^2_g$
powers gives
\begin{equation}
V(q^2) = - \frac{4}{3}\; \frac{4\pi}{q^2}\;\biggl(\frac{8\pi}{\beta_0}\;
\frac{\mu^2_g}{q^2} + \frac{4\pi}{\beta_0} - \frac{\alpha_s(\mu^2_g)}{2}\biggr)
+ O(1)\;, \label{ev}
\end{equation}
so that at large $|{\bf x}|$ values and at $\beta_0(n_f=3) = 9$ one gets
\begin{equation}
V({\bf x}) = \frac{16\pi}{27}\; \mu^2_g |{\bf x}| - \frac{4}{3}\;
\frac{1}{|{\bf x}|}\biggl(\frac{4\pi}{9}-\frac{\alpha_s(\mu^2_g)}{2}\biggr)
+o(1/|{\bf x}|)\;. \label{eqv}
\end{equation}
Note, that the coefficient at the scale, giving the $\sigma|{\bf x}|$ term
in the potential, is twice greater, than in eq.(\ref{8-}). If one accepts the
given scheme of matching of the gluon propagator parameters with the parameters
of potential, confining the quarks in the heavy quarkonium, then at
$\mu^2 = \mu^2_g(1+O(q^2/\mu^2_g))$, one will have
$$
\Lambda(\mu) \simeq \Lambda(\mu_g) = \mu_g\;,
$$
and we obtain, that the effective "running" coupling constant of QCD is given
in the $\overline{V}$-scheme. Moreover, the $\Lambda_{lin}$ value, determined
in
Section 2 in the $1/n_f$ consideration, is related with the evolution parameter
of the coupling constant by the expression
$$
\frac{\Lambda_{lin}}{\mu_g} = \sqrt{2}\;,
$$
that is in a good agreement with the experimental value in eq.(\ref{e2}).
Thus, the matching of the evolution scale $\mu_g$ with the gluon string tension
in the heavy quarkonium results in the value
\begin{equation}
\mu_g^{\overline{\rm MS}} = 136\pm 4\; \mbox{MeV} \label{mg}
\end{equation}
at $n_f=3$.

Further, as was shown, the account for the fermion loop contribution into the
gluon propagator gives the potential of the Richardson's form, with
$\Lambda(\mu) = \mu_g$. In such potential, the expansion over the $q^2/\mu^2_g$
powers gives
\begin{equation}
V^{\rm Rich}(q^2) = - \frac{4}{3}\; \frac{4\pi}{q^2}\;
\biggl(\frac{4\pi}{\beta_0}\; \frac{\mu^2_g}{q^2} + \frac{2\pi}{\beta_0}\biggr)
+O(1)\;. \label{er}
\end{equation}
Comparing eq.(\ref{ev}) and eq.(\ref{er}), one notes, that the modified
potential (\ref{eqv}) at large distances can be obtained from the
Richardson's potential by the following
\begin{equation}
V({\bf x})\approx \frac{8\pi}{27}\;\mu_g^2 |{\bf x}| + V^{\rm Rich}({\bf x})\;,
\label{vm}
\end{equation}
if one supposes
$$
\alpha_s(\mu^2_g) = \frac{4\pi}{\beta_0}\;,
$$
i.e.
\begin{equation}
\ln\frac{\mu_g^2}{\Lambda^2_{QCD}} = 1\;.\label{ml}
\end{equation}
The additional term in eq.(\ref{vm}) appears as the correction to the linear
potential, confining the quarks, and its presence is due to the modification
of the renormalon. Furthermore, the matching of the modified potential with
the Richardson's potential over the coulomb deviation from the linear rise at
large $|{\bf x}|$ allows one to fix the ratio $\mu_g/\Lambda_{QCD}$ or
$\alpha_s(\mu_g^2)$ (see eq.(\ref{ml})). As has to be expected, the presence
of the additional scale in the $\alpha_s D$ quantity becomes unessential at low
distances, i.e. at large momentum transfers, when the dependence on the scale
is determined by the coupling constant evolution.
\setlength{\unitlength}{1.5mm}
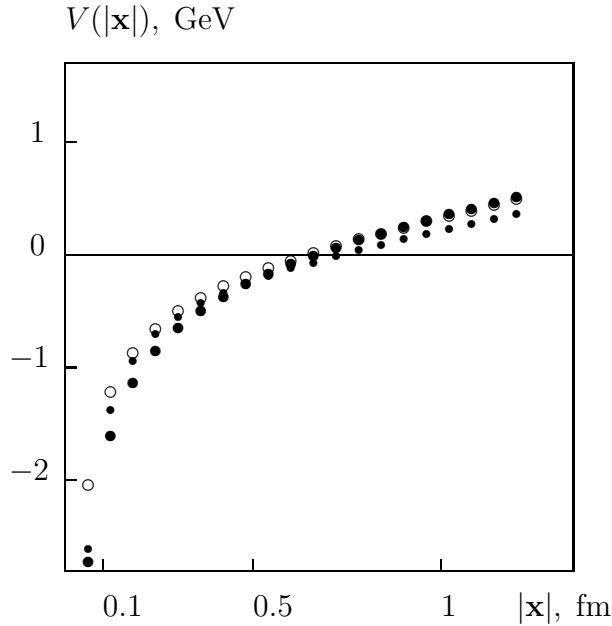
\begin{figure}[t]
\begin{center}
\begin{picture}(50,50)
\put(0,2){\framebox(45,45)}
\put(0,30){\line(1,0){45}}
\put(3.33,2){\line(0,1){1}}
\put(16.66,2){\line(0,1){1}}
\put(33.3,2){\line(0,1){1}}
\put(3.33,-2){$0.1$}
\put(16.66,-2){$0.5$}
\put(33.3,-2){$1$}
\put(40,-2){$|{\bf x}|,\; \mbox{fm}$}
\put(0,20){\line(1,0){1}}
\put(0,10){\line(1,0){1}}
\put(0,40){\line(1,0){1}}
\put(-5,10){$-2$}
\put(-5,20){$-1$}
\put(-5,40){$~~1$}
\put(-5,30){$~~0$}
\put(0,50){$V(|{\bf x}|),\; \mbox{GeV}$}
 \put(  2.00,  2.80){\circle*{1}}
 \put(  2.00,  9.60){\circle{1}}
 \put(  2.00,  3.90){\circle*{0.7}}
 \put(  4.00, 13.90){\circle*{1}}
 \put(  4.00, 17.80){\circle{1}}
 \put(  4.00, 16.20){\circle*{0.7}}
 \put(  6.00, 18.60){\circle*{1}}
 \put(  6.00, 21.30){\circle{1}}
 \put(  6.00, 20.60){\circle*{0.7}}
 \put(  8.00, 21.50){\circle*{1}}
 \put(  8.00, 23.40){\circle{1}}
 \put(  8.00, 23.00){\circle*{0.7}}
 \put( 10.00, 23.50){\circle*{1}}
 \put( 10.00, 25.00){\circle{1}}
 \put( 10.00, 24.50){\circle*{0.7}}
 \put( 12.00, 25.00){\circle*{1}}
 \put( 12.00, 26.20){\circle{1}}
 \put( 12.00, 25.70){\circle*{0.7}}
 \put( 14.00, 26.30){\circle*{1}}
 \put( 14.00, 27.20){\circle{1}}
 \put( 14.00, 26.60){\circle*{0.7}}
 \put( 16.00, 27.40){\circle*{1}}
 \put( 16.00, 28.00){\circle{1}}
 \put( 16.00, 27.40){\circle*{0.7}}
 \put( 18.00, 28.30){\circle*{1}}
 \put( 18.00, 28.80){\circle{1}}
 \put( 18.00, 28.10){\circle*{0.7}}
 \put( 20.00, 29.20){\circle*{1}}
 \put( 20.00, 29.50){\circle{1}}
 \put( 20.00, 28.80){\circle*{0.7}}
 \put( 22.00, 29.90){\circle*{1}}
 \put( 22.00, 30.20){\circle{1}}
 \put( 22.00, 29.30){\circle*{0.7}}
 \put( 24.00, 30.60){\circle*{1}}
 \put( 24.00, 30.80){\circle{1}}
 \put( 24.00, 29.90){\circle*{0.7}}
 \put( 26.00, 31.30){\circle*{1}}
 \put( 26.00, 31.40){\circle{1}}
 \put( 26.00, 30.40){\circle*{0.7}}
 \put( 28.00, 31.90){\circle*{1}}
 \put( 28.00, 31.90){\circle{1}}
 \put( 28.00, 30.90){\circle*{0.7}}
 \put( 30.00, 32.50){\circle*{1}}
 \put( 30.00, 32.40){\circle{1}}
 \put( 30.00, 31.40){\circle*{0.7}}
 \put( 32.00, 33.00){\circle*{1}}
 \put( 32.00, 33.00){\circle{1}}
 \put( 32.00, 31.80){\circle*{0.7}}
 \put( 34.00, 33.60){\circle*{1}}
 \put( 34.00, 33.40){\circle{1}}
 \put( 34.00, 32.30){\circle*{0.7}}
 \put( 36.00, 34.10){\circle*{1}}
 \put( 36.00, 33.90){\circle{1}}
 \put( 36.00, 32.70){\circle*{0.7}}
 \put( 38.00, 34.60){\circle*{1}}
 \put( 38.00, 34.40){\circle{1}}
 \put( 38.00, 33.20){\circle*{0.7}}
 \put( 40.00, 35.10){\circle*{1}}
 \put( 40.00, 34.90){\circle{1}}
 \put( 40.00, 33.60){\circle*{0.7}}

\end{picture}
\end{center}
\caption{The potentials, corresponding to the Cornell model (dots),
the model with the "running" $\alpha_s/\omega$ value (solid circles) and with
the "running" $\bar \alpha_s$ value plus the linear correction (empty circles)}
\end{figure}

The comparison of the modified potential with eq.(\ref{vm}) and with the
Cornell model potential is given on figure 1. As one can see, at large
$|{\bf x}|> 0.5$ fm, where the modification is valid, the Richardson's
potential
with $\mu_g$, determined in eq.(\ref{mg}), and with the account for the
additional linear term is the quite accurate approximation, that is also close
to the potential of Cornell model. As was noted \cite{eich}, the QCD-motivated
potentials, fitting the spectra of charmonium and bottomonium, give one and
the same form of the $V({\bf x})$ dependence at $0.2 <|{\bf x}|<1$ fm with
the accuracy up to an additive shift $\Delta V$, independent of $|{\bf x}|$.
This form is caused by the transition of the coulomb-like potential, following
from the asymptotic freedom of QCD at low distances, to the linearly rising
string-like potential, confining the quarks at large distances $|{\bf x}|\ge
1$ fm. In the present section we have found the connection between the
scale of the QCD "running" constant evolution and the string tension.

\section{Model-dependent estimate of scales}

A toy ansatz, based on the extraction of the pole singularity of $\omega$ in
the gluon propagator in the euclidian space and in the $V$-scheme is given by
the expression
$$
D_{mod}(k^2) = \frac{4\ln(\mu_g/\Lambda_{QCD})}{k^2-\mu_g^2}\;,
$$
that is used for the one-loop calculation of the gluon  condensate
$\langle\frac{\alpha_s}{\pi} G^2\rangle$ value. To get the renormalization
group invariant expression in the dimensional regularization, one has to
extend $D_{mod}$ into the $(4-2\epsilon)$-dimensional space, so that
$D_{mod}^{(4-2\epsilon)} = D_{mod}(1+\epsilon z_1)$, where $z_1$ is a
single-fold determined value, because of the requirement of the renormalization
group invariance of the gluon condensate value. Straightforward calculations
give
\begin{equation}
\big\langle \frac{\alpha_s}{\pi} G^2\big\rangle = \frac{32}{\pi^2}
\ln\frac{\mu_g}{\Lambda_{QCD}}\;\biggl(\ln\frac{\mu_g}{\Lambda_{QCD}}+
\frac{1}{6}\biggr)\;\mu_g^4\;.\label{cond}
\end{equation}
Note, that in accordance with eq.(\ref{cond}), the gluon condensate tends to
zero, if one has the purely perturbative-like expression for $D_{mod}$
at $\mu_g=0$ or if one takes the fixed point $\mu_g=\Lambda_{QCD}$.
The substitution of numbers for the scales, estimated in the previous
section, results in
\begin{equation}
\big\langle \frac{\alpha_s}{\pi} G^2\big\rangle \approx 0.011\;
\mbox{GeV}^4\;, \label{num}
\end{equation}
where $\mu_g = e^{5/6}\mu_g^{\overline{\rm MS}} = 316\pm 12$ MeV. Value
(\ref{num}) is in a good agreement with the gluon condensate estimate in
the framework of QCD sum rules \cite{sr}.

\section*{Conclusion}

In the present paper we have shown, that in the $1/n_f$ consideration, the
account for the anomalous scale dependence of the gluon propagator  leads
to the modification of the renormalon, because of the redefinition of the
evolution scale for the "running" coupling constants of QCD. When one restores
the correct renormalization group dependence of the gluon
propagator on the scale in the nonabelian theory, the modified renormalon
obtains the hidden scale dependence, that is expressed in the specific
transformation properties of the effective "running" coupling constant. The
phenomenological matching of the modified renormalon with the tension of gluon
string between the static quarks and with the subleading coulomb  interaction
at large distances results in the fixing of the two renormalization group
invariant scales of the modified renormalon, and it gives the normalization
condition for the "running" coupling constant.
The modification of renormalon leads to the linear term, additional to the
Richardson's potential. Furthermore, we have made the model estimate of the
gluon condensate value at the presence of two infrared parameters,
determining the scales of the pole singularities in the "running" QCD constant
and in the gluon propagator up to one loop. At large virtualities of the gluon,
the coupling constant is given by the expression
$$
\bar \alpha_s(m^2,\mu_g) = \frac{2\pi}{\beta_0\ln(m/\mu_g)}\;,
$$
to one loop, so that
$$
\mu_g^{\overline{\rm MS}}(n_f=3) = 136\pm 4\; \mbox{MeV,}
$$
that is in  agreement with the experimental estimates of $\alpha_s(m_Z^2)$.
Moreover,
$$
\alpha_s(\mu_g) = \frac{2\pi}{\beta_0\ln(\mu_g/\Lambda_{QCD})}
= \frac{4\pi}{9}\;.
$$

This work is partially supported by the International Science Foundation
grants NJQ000, NJQ300 and by the program "Russian State Stipendia for young
scientists".

\vspace*{0.4cm}

\hfill {\it Received June 9, 1995}
\end{document}